\documentclass[conference]{IEEEtran}
\ifCLASSINFOpdf
\else
\fi
\usepackage{amsmath}
\usepackage{array}

\usepackage{graphicx}
\usepackage{booktabs}
\usepackage[caption=false]{subfig}
\usepackage{multirow}
\IEEEoverridecommandlockouts

\begin{document}
%
\title{Evaluation of preference of multimedia content using deep neural networks for electroencephalography}

\author{\IEEEauthorblockN{Seong-Eun Moon \qquad Soobeom Jang \qquad Jong-Seok Lee \thanks{\copyright 2018 IEEE. Personal use of this material is permitted. Permission from IEEE must be obtained for all other uses, in any current or future media, including reprinting/republishing this material for advertising or promotional purposes, creating new collective works, for resale or redistribution to servers or lists, or reuse of any copyrighted component of this work in other works.}}
\IEEEauthorblockA{School of Integrated Technology \\ Yonsei University\\
Republic of Korea\\
Email: \{se.moon, soobeom.jang, jong-seok.lee\}@yonsei.ac.kr}
}


%


\maketitle

\begin{abstract}
Evaluation of quality of experience (QoE) based on electroencephalography (EEG) has received great attention due to its capability of real-time QoE monitoring of users. However, it still suffers from rather low recognition accuracy. In this paper, we propose a novel method using deep neural networks toward improved modeling of EEG and thereby improved recognition accuracy. In particular, we aim to model spatio-temporal characteristics relevant for QoE analysis within learning models. The results demonstrate the effectiveness of the proposed method.
\end{abstract}


%
\IEEEpeerreviewmaketitle

\section{Introduction}
\label{sec:intro}

With the noticeable growing of the demand for multimedia content, user-adaptive content delivery has become a key to success of many multimedia services. Consequently, it is crucial to understand how users perceive the multimedia, which is the concept of quality of experience (QoE) defined as ``the degree of delight or annoyance of the user of an application or service'' \cite{Qualinet13}. 
Traditionally, QoE has been measured explicitly, i.e., subjects are asked about their experience with the given multimedia content via an interview or a questionnaire. However, it is difficult for this explicit approach to capture the user responses in real-time because the evaluation is typically implemented after the presentation of content. 

On the other hand, QoE also can be monitored through the implicit cues obtained from the physiological or behavioral signals of users, which enables real-time monitoring of QoE. Particularly, the brain signals such as electroencephalography (EEG) are expected to provide deeper insight into the perceptual experience of multimedia because they contain the whole information of the multimedia perception, whereas the explicit approach can measure only predefined final outputs of the perception.

Many studies have employed the EEG signals to capture the degradation of QoE \cite{Antons12, Scholler12, Mustafa12} and the overall QoE \cite{Antons13,Arndt14,Koelstra12,Moon15,Perrin15,Kroupi14,Kroupi16}. They showed the potential of EEG to automatically monitor QoE of users, which can be used for many applications such as QoE-aware video scaling for content delivery \cite{Lee12} and personalized multimedia recommendation \cite{Moon17survey}. 

A limitation of the existing EEG-based implicit QoE assessment systems is that their performance still remains at insufficient levels for real-world applications where high reliability is critical. For instance, Table \ref{table:previous} summarizes the results reported in representative studies on binary classification of content preference for the DEAP database \cite{Koelstra12}. It can be seen that even the recent deep learning approaches show accuracies lower than 90\%.

\begin{table}
	\renewcommand{\arraystretch}{1.5}
	\centering
	\caption{Previous results of the binary classification of preference for the DEAP database.}~\label{table:previous}
	\begin{tabular}{>{\centering}m{0.5cm} >{\centering}m{2cm} >{\centering}m{2.5cm} >{\centering}m{1.5cm}} \toprule[1pt]
		\textbf{Ref.}		& \textbf{Classifier} 				& \textbf{Classification scheme} & \textbf{Classification accuracy} \tabularnewline \toprule[1pt]
		
		\cite{Koelstra12}	& Gaussian naive Bayes classifier	& leave-one-video-out for each subject	& 0.502 (F1-score) \tabularnewline 
		
		\cite{Gupta16}		& Relevance vector machine			& leave-one-video-out for each subject & 0.65 (F1-score) \tabularnewline 
		
		\cite{Amjadzadeh17}	& Ensemble classifier* & leave-one-trial-out & 0.647 \tabularnewline 
		
		\cite{Zhuang14}		& Support vector machine			& leave-one-video-out for each subject & 0.705 \tabularnewline 
		
		\cite{Xu16}			& Deep belief network				& five-fold cross-validation for each subject & 0.867 (F1-score) \tabularnewline 
		
		\cite{Alhagry17}	& Recurrent neural network			& four-fold cross-validation & 0.880 \tabularnewline \bottomrule
		
	\end{tabular}
	\\
	\flushleft
	*Ensemble of support vector machine, nearest mean, 1-nearest neighbor, k-nearest neighbor, and linear discriminant analysis
\end{table}

We notice that the spatial relationship of EEG signals has not been significantly considered in the previous EEG-based QoE recognition studies although it possibly includes useful information of neural activities. 
In the resting state of the brain, the neural activities of different brain regions show a certain relationship that comprises the functional resting-state network \cite{Beckmann05}. However, if any stimulus is given, the spatial relationship is altered because the neural activity of interest appears. 

In this paper, we propose a novel approach to improve the accuracy of EEG-based preference recognition. Particularly, convolutional neural networks (CNNs) are employed, which has the capability to analyze the spatial information of EEG signals. 
The contributions of this work can be summarized as follows:
\begin{itemize}
	\item We achieve high recognition accuracy of preference based on EEG by adopting deep CNNs that enable spatial analysis of EEG signals. This demonstrates the feasibility of the real-world applications using EEG such as real-time QoE monitoring, automatic feedback generation, QoE-aware multimedia compression, and so on.
	\item We compare various types of EEG features, input shapes for CNNs, and CNN structures with different complexity, which contributes to further related studies by providing guidelines for system design.  
\end{itemize}

\section{EEG signal features}
\label{sec:feature}
This section describes the EEG features that are employed as inputs of the CNNs. They can be categorized depending on whether the feature considers the activation of a single region or multiple regions. In this paper, one feature indicating the activation level of a single brain region and three features that consider the activation of multiple brain regions are employed. Details of the features are explained below.

\subsection{Power spectral density (PSD)}
\label{subsec:psd}
PSD represents the activation level of a single electrode. It is calculated using the Welch's method, which is a non-parametric spectral estimation method based on the Fourier transform. For the $i$-th window of EEG signals $x_i, (i=1, 2, ..., M)$, the periodogram at frequency $f$ is calculated as:
\begin{equation}
P_i(f) = \frac{1}{NW}\left| \sum_{n=1}^{N} w(n)x_i(n)e^{-j2\pi fn} \right| ^2,
\end{equation}
where $N$ is the number of data points in the window, $w(n)$ is a window function, and $W$ is a normalization constant given as $W=\frac{1}{N}\sum_{n=1}^{N} \left|w(n)\right|^2$. PSD is obtained by averaging the periodogram over the windows:
\begin{equation}
	PSD(f) = \frac{1}{M}\sum_{i=1}^{M} P_i(f).
\end{equation}

Furthermore, the PSD values of the baseline signals (five seconds before presentation of stimuli) are subtracted from those of the corresponding trial signals to eliminate irrelevant neural activities.

\subsection{Pearson correlation coefficient (PCC)}
\label{subsec:pcc}
PCC is a measure of the linear relationship between two signals, which ranges from -1 to 1. A PCC value of -1 (1) corresponds to the perfect negative (positive) linear relationship, and a PCC value of zero indicates that there is no linear relationship between the two signals. It is calculated as follows:
\begin{equation}
	PCC = \frac{cov(X,Y)}{\sigma_X\sigma_Y},
\end{equation}
where $\sigma_X$ and $\sigma_Y$ indicate the standard deviations of the given two signals $X=\{x_i\}$ and $Y=\{y_i\}$, respectively, and $cov(X,Y)$ is the covariance between them. 

\subsection{Phase locking value (PLV)}
\label{subsec:plv}
PLV \cite{Lachaux99} describes the phase synchronization between two signals, which is calculated as an absolute value of the average phase differences over temporal windows. This can be presented as:
\begin{equation}
	PLV = \frac{1}{M} \left| \sum_{i=1}^{M} e^{j\Delta\phi_i} \right|,
\end{equation}
where $\Delta\phi_i$ indicates the phase difference of the $i$-th window. PLV ranges between 0 to 1, which correspond to independence and perfect synchronization of two signals, respectively.

\subsection{Transfer entropy (TE)}
\label{subsec:te}
TE \cite{Schreiber00} measures information flow between two time series, assuming that the two time series can be approximated by Markov chains. It is defined as:
\begin{equation}
	TE_{Y\to X} = \sum p(x_{i+1},x_i,y_i)\log \frac{p(x_{i+1}|x_i,y_i)}{p(x_{i+1}|x_i)}.
	\label{eq:te}
\end{equation}
The result of (\ref{eq:te}) is the directional information that indicates the ability of time series $Y$ to improve the prediction of time series $X$. 
We use the Java Information Dynamics Toolkit \cite{Lizier14} to obtain TE features.

\section{System design}
\label{sec:system}

\subsection{Input}
\label{subsec:input}

\begin{figure}[t]
	\centering
	\subfloat[theta]{\includegraphics[width=0.35\columnwidth]{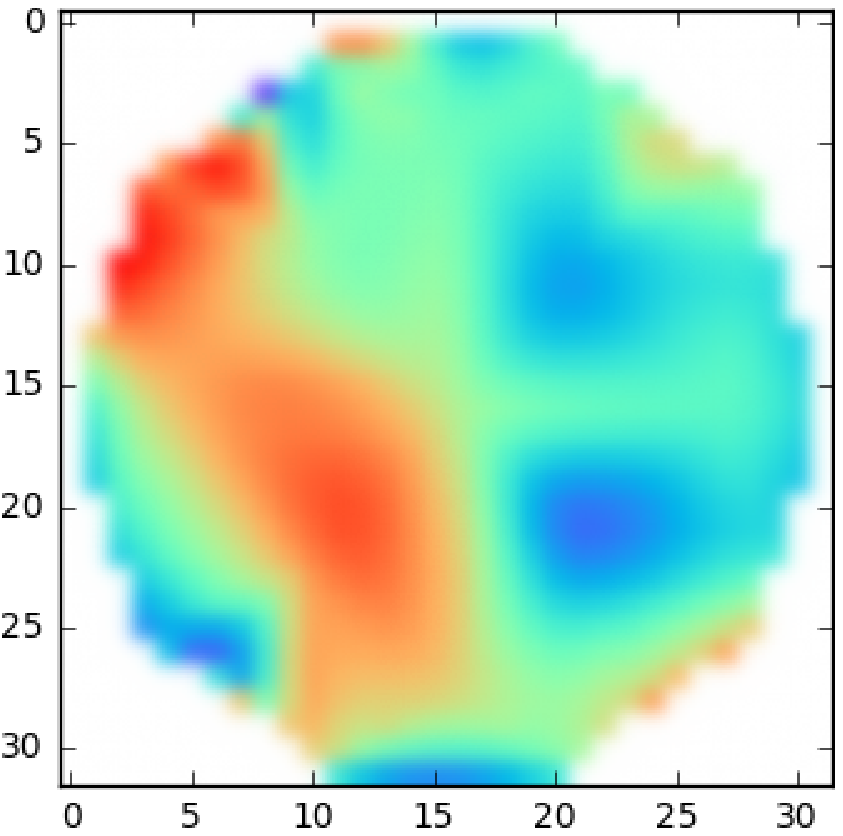}} \quad
	\subfloat[alpha]{\includegraphics[width=0.35\columnwidth]{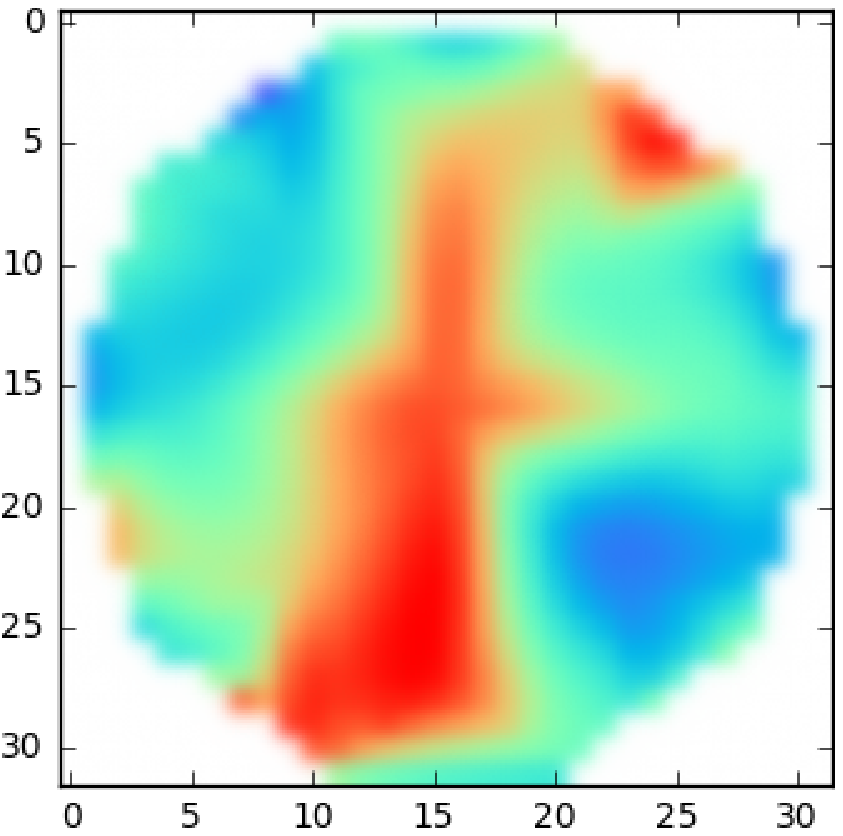}} \hfill
	\subfloat[beta]{\includegraphics[width=0.35\columnwidth]{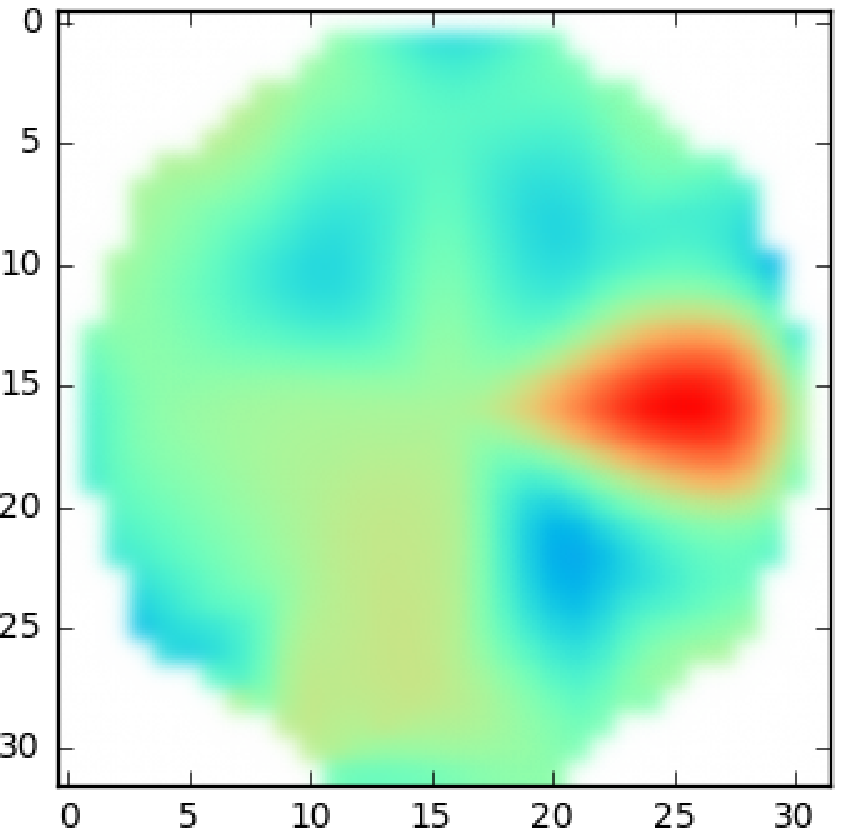}} \quad
	\subfloat[gamma]{\includegraphics[width=0.35\columnwidth]{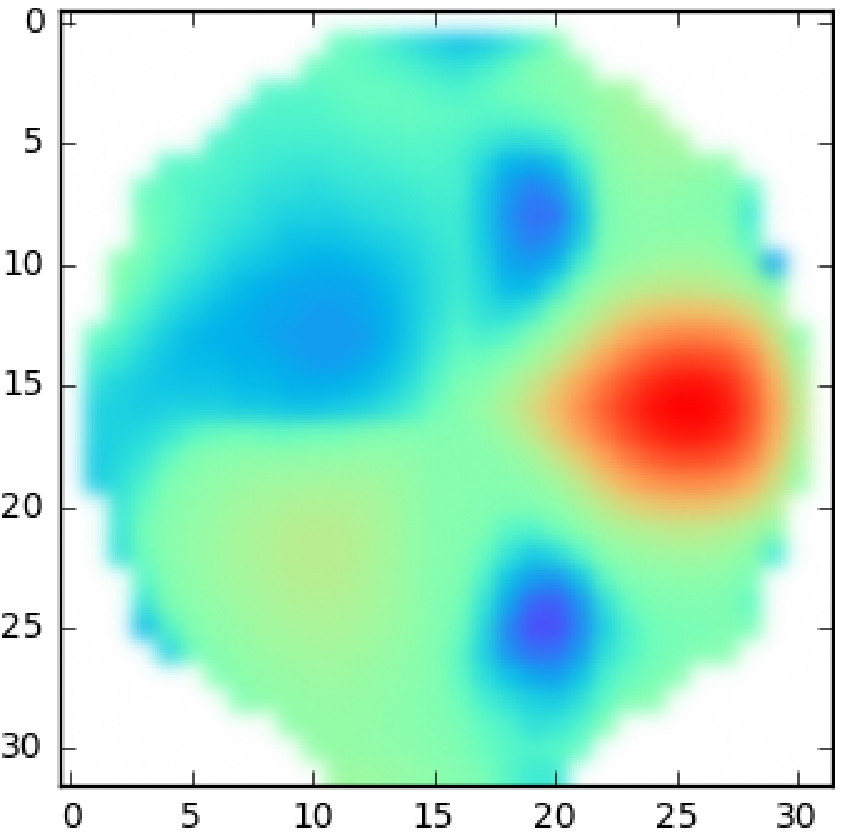}}
	\caption{Examples of PSD topographies for different frequency bands. A region marked by a red (or blue) color indicates the most positively (or negatively) activated EEG signals, and a region marked with a green color corresponds to a zero activation compared with the baseline signals. }
	\label{fig:topoexp}
\end{figure}
PSD indicates the activation level of a single regions of the brain. Therefore, the PSD values can be represented as a topography, i.e., they are allocated to the locations of the corresponding electrodes and the rest of the scalp is filled by interpolation \cite{Bashivan16}. Examples of the topography are shown in Figure \ref{fig:topoexp}. The outside of the head is filled with zeros. 

In contrast to PSD, it is difficult to describe the other features as a topographic figure because they measure the relationship between two regions of the brain, which is called the brain connectivity \cite{Friston11}. We transform the features into matrices used as CNN inputs, whose ($i, j$)-th element is the feature value obtained by using the data of the $i$-th and $j$-th electrodes. 

\begin{figure}[!t]
	\centering
	\includegraphics[width=0.6\columnwidth]{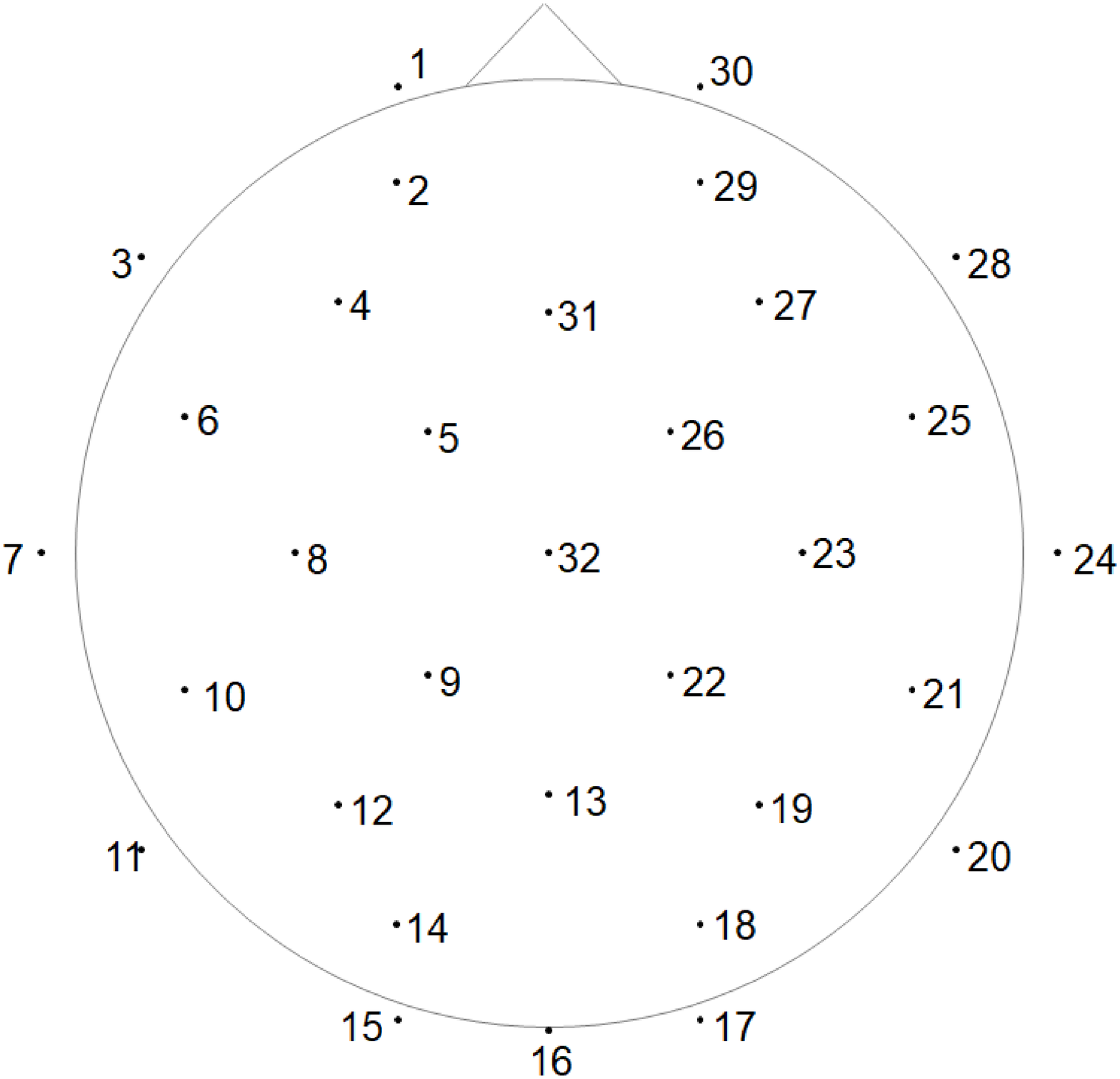}
	\caption{Ordering method of the EEG electrodes based on the distance between electrodes and hemispheric structure of the head.}
	\label{fig:dist1}
\end{figure}

Here, the order of electrodes in the input matrix becomes important because the filters of a CNN learn localized patterns of the matrix. We consider two different ordering methods, namely, `distance' and `random'. The first method arranges the EEG electrodes according to the distance between two electrodes so that physically neighboring electrodes are adjacent in the matrix. At the same time, it considers the hemispheric structure of the brain as shown in Figure \ref{fig:dist1}. That is, the ordering starts from the left frontal electrode and proceeds to the nearest electrode in the depth direction of the head within the left hemisphere; after finishing the left side of the head, it shifts to the occipital area of the right hemisphere, repeats the same process for the right side, and ends at the center. The second method simply randomizes the order of electrodes. 

\subsection{CNN structure}
\label{subsec:cnn}
\begin{figure}[!t]
	\centering
	\includegraphics[width=\columnwidth]{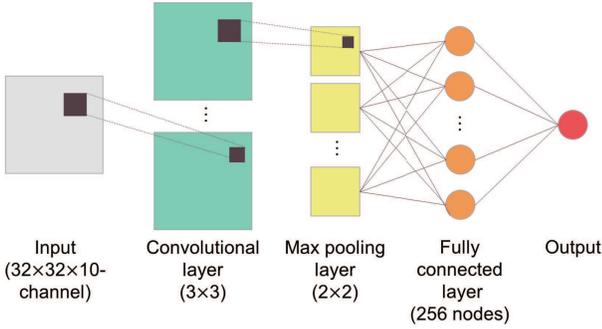}
	\caption{Example CNN structure.}
	\label{fig:cnn}
\end{figure}
Three CNN structures with different complexity are adopted for the EEG-based preference recognition. The simplest structure includes one convolutional layer and one max-pooling layer as illustrated in Figure \ref{fig:cnn}. The second structure has one convolutional layer, one max-pooling layer, two convolutional layers, one max-pooling layer, and finally a fully connected layer. The third structure consists of five convolutional layers and five max-pooling layers, one after the other, before the fully connected layer. The three CNN structures are denoted as CNN1, CNN2, and CNN3 in the following.

The first convolutional layer of each CNN structure has 32 filters, and the number of filters of the following convolutional layer becomes twice that of the previous convolutional layer. The size of the filters is fixed as 3$\times$3 for all convolutional layers. The rectified linear unit (ReLU) is employed as the activation function. The max-pooling is conducted for 2$\times$2 patches, and the batch normalization is implemented after every max-pooling.

The CNNs are implemented in Theano. The Adam algorithm is used for training by minimizing the loss defined by the cross-entropy function. The training is conducted with a Tesla K80 GPU, where the batch size is set to 256.

\subsection{Database}
\label{subsec:data}
We employ the DEAP database \cite{Koelstra12} that contains one-minute-long 32-channel EEG signals recorded while 32 subjects were watching videos and the corresponding preference scores that indicate how much the subjects like the videos. It has been popularly used for analyzing multimedia experience based on EEG.
As the number of electrodes is 32, the sizes of the feature matrices for PCC, PLV, and TE become 32$\times$32, and the topographies for PSD are also rendered into 32$\times$32 pixels to have the same input size.

The EEG signals are divided into three-second-long segments with an overlap of 2.5 seconds. Thus, the total number of data is 147,200 (32 subjects$\times$40 videos$\times$115 segments). These data are divided into five clusters randomly, which are used for a five-fold leave-one-cluster-out cross-validation scheme. 
The features are calculated for delta (0-3 Hz), theta (4-7 Hz), low alpha (8-9.5 Hz), high alpha (10.5-12 Hz), alpha (8-12 Hz), low beta (13-16 Hz), mid beta (17-20 Hz), high beta (21-29 Hz), beta (13-29 Hz), and gamma (30-50 Hz) frequency bands of EEG signals. Consequently, the sizes of CNN inputs become 32$\times$32$\times$10.

We examine two scenarios of preference prediction. First, a binary classification of preference (liking vs. disliking) is considered. As the original preference score in the database lies in a 9-point scoring scale, we define the videos received preference scores between 1 and 5 as one class, and the rest as the other. As a result, 33.52\% of the entire data are labeled as the `disliking' class, and 66.48\% of the data are assigned as the `liking' class. Note that the sizes of the two classes are highly imbalanced. Therefore, the F1-score is used for evaluation of prediction results. For the random ordering, the final result is obtained by averaging the F1-score with three differently randomized orders.
Second, the subjective preference score is estimated, which is a regression task. The regression performance is assessed in terms of the root-mean-square error (RMSE) between the ground truth and predicted preference scores.

\section{Results}
\label{sec:res}

\begin{table}[!t]
	\renewcommand{\arraystretch}{1.3}
	\centering
	\caption{F1-scores of binary preference classification.}~\label{table:clsresult}
	\begin{tabular}{>{\centering}m{1.5cm} c c c c c} \toprule[1pt]
		& & \multicolumn{4}{c}{\textbf{Feature type}} \\ \cmidrule{3-6}
		  & & PCC & PLV & TE & PSD \\ \toprule[1pt]
		\multirow{2}{*}{CNN1} & distance & 0.932 & 0.967 & 0.945 & \multirow{2}{*}{0.762} \\
							   & random   & 0.936 & 0.966 & 0.942 & \\ \midrule
		\multirow{2}{*}{CNN2} & distance & 0.938 & 0.969 & 0.921 & \multirow{2}{*}{0.791} \\
							   & random   & 0.937 & 0.959 & 0.911 & \\ \midrule
		\multirow{2}{*}{CNN3}& distance & 0.927 & 0.907 & 0.811 & \multirow{2}{*}{0.814} \\
							   & random   & 0.914 & 0.895 & 0.808 & \\ \bottomrule		
	\end{tabular}
\end{table}

\subsection{Binary classification}
\label{subsec:res-cls}
The results of binary classification are shown in Table \ref{table:clsresult}. Overall, the proposed system results in much higher F1-scores than the previous works shown in Table \ref{table:previous}. The best performance (F1-score = 0.969) is obtained with the combination of CNN2, the distance-based ordering method, and PLV. There are also several other cases showing comparable performance to the best case.

When the complexity of CNNs is examined, the obtained results indicate that a more complex structure does not necessarily produces better classification performance. The most complex network, i.e., CNN3, show rather degraded classification performance compared with the simpler networks for PCC, PLV, and TE, and CNN2 also results in lower F1-scores for TE and PLV (random order). Only the performance of PSD is improved by adopting more complex CNN architectures. 

The features concerning the relationship between different brain regions demonstrate better classification results, i.e., F1-scores with PCC, PLV, and TE significantly exceed those with PSD, except for the case with TE and CNN3. In particular, PLV shows the best performance among the features with relatively shallow CNN structures (CNN1 and CNN2), but PCC is better than the other features with the most complex CNN structure (CNN3). 

The prediction performance also varies depending on the ordering method. Overall, the physical distance-based ordering yields better accuracy than the random ordering, which indicates the strategy of distance-based ordering, highlighting the information of interest by allocating a single receptive field to feature values that are possibly similar to each other, works better for recognition.

\begin{table}[!t]
	\renewcommand{\arraystretch}{1.3}
	\centering
	\caption{RMSEs of preference score regression.}~\label{table:regresult}
	\begin{tabular}{>{\centering}m{1.5cm} c c c c c} \toprule[1pt]
		& & \multicolumn{4}{c}{\textbf{Feature type}} \\ \cmidrule{3-6}
		& & PCC & PLV & TE & PSD \\ \toprule[1pt]
		\multirow{2}{*}{CNN1} & distance & 1.538 & 1.429 & 1.557 & \multirow{2}{*}{2.064} \\
		& random   & 1.536 & 1.417 & 1.546 & \\ \midrule
		\multirow{2}{*}{CNN2} & distance & 1.340 & 1.320 & 1.517 & \multirow{2}{*}{1.858} \\
		& random   & 1.386 & 1.309 & 1.549 & \\ \midrule
		\multirow{2}{*}{CNN3}& distance & 1.238 & 1.228 & 1.643 & \multirow{2}{*}{1.741} \\
		& random   & 1.252 & 1.237 & 1.644 & \\ \bottomrule		
	\end{tabular}
\end{table}
 
\subsection{Score regression}
\label{subsec:res-reg}

Table \ref{table:regresult} shows the results of the preference score regression. The best result (RMSE = 1.228) is achieved when PLV matrices formed using the distance-based method are employed for CNN3. This demonstrates that it is feasible to specify the level of preference in a finer scale than the binary classification. 

As in the binary classification results, the features containing relational information outperform PSD. The RMSE of PSD is always larger than those of PCC, PLV, and TE, and PLV shows the best performance for all CNN structures.

The influence of the complexity of CNN structures is different from that for the classification. The results of PCC, PLV, and PSD are improved by using more complex networks; for TE, the best performance is obtained with CNN2. 

Furthermore, although the distance-based ordering consistently provides better performance for CNN3, the result of the random ordering is better, in particular, for CNN1.

\subsection{Discussion}
\label{subsec:res-dis}

From the results of the binary classification and score regression, we consistently observed the superiority of the features that measure the relationship between different brain regions. That is, such relationship includes useful information for prediction of preference. 

In particular, we compared three features that reflect different aspects of the relationship. PLV that measures the phase synchronization between brain regions achieves high prediction accuracy with relatively shallow structures in the binary classification. In the score regression, PLV consistently provides the best performance regardless of CNN structures. 

PCC and PLV are rather traditional, simple approaches to analyze the brain connectivity without consideration of directionality, but show better performance than TE measuring directional information flow between different regions. This is somewhat surprising, and would require further investigation in the future. 

The influence of the network complexity was also examined from the results. In the binary classification, the CNN structure with medium complexity provides the best recognition result, and the most complex network shows the worse result. However, a more complex network shows a better result overall for the score regression. This is probably because the score regression is more difficult to solve than the binary classification, i.e., the mapping function of the score regression is more complex than that of the binary classification.

It is observed that feature matrices aligned according to the distance-based order performs better than those using the random order when the deep CNN structure is used, whereas such superiority is not prominent (binary classification) or not observed (score regression) in CNN1. 
This is probably because the structure of CNN1 is not complex enough to take the advantage of distance-based ordering.

\begin{figure}[!t]
	\centering
	\subfloat[subject]{\includegraphics[width=\columnwidth]{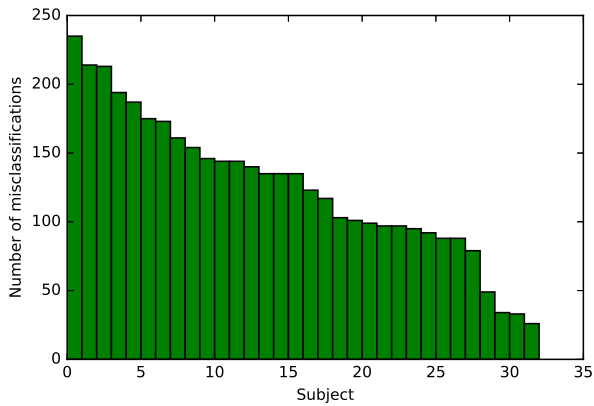}} \hfill
	\subfloat[video]{\includegraphics[width=\columnwidth]{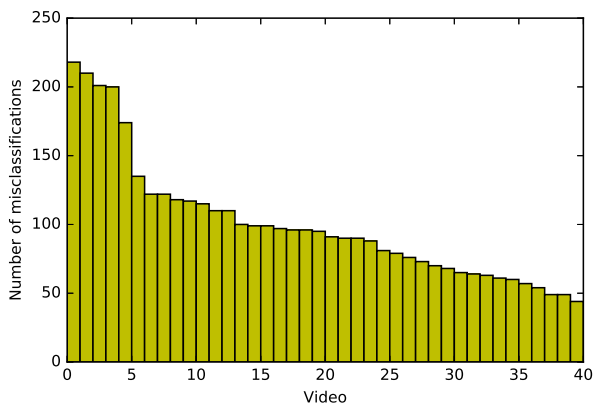}}
	\caption{Histograms of misclassifications with respect to the (a) subject and (b) video.}
	\label{fig:fail}
\end{figure}

We further analyze the failure cases of the binary classification to verify whether such cases are influenced by the specific subject or video. Histograms of the number of misclassifications for the best case in Table \ref{table:clsresult} are shown in Figure \ref{fig:fail}. The indexes of subjects and videos are in a descending order of the number of failures. 

It can be noticed that the recognition performance notably varies depending on the subject. The first three subjects take 17\% of the failure cases, whereas the last three subjects occupy only 3\%. As no noticeable rating tendency is found for those subjects, this variance of recognition performance indicates that there is significant individual difference in neural activities related to QoE. 

\begin{figure}[!t]
	\centering
	\subfloat[preference]{\includegraphics[width=0.9\columnwidth]{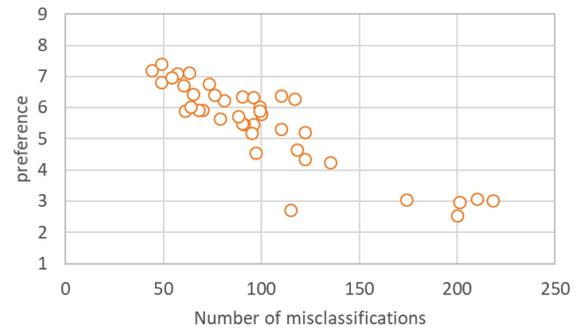}} \hfill
	\subfloat[valence]{\includegraphics[width=0.9\columnwidth]{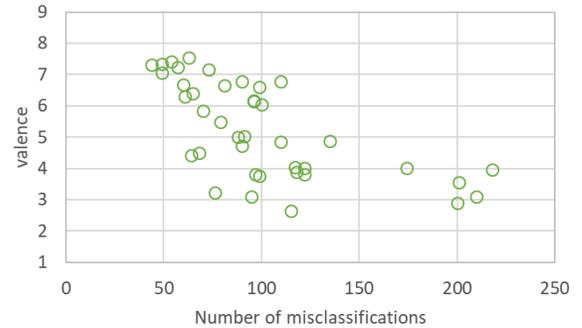}} \hfill
	\subfloat[arousal]{\includegraphics[width=0.9\columnwidth]{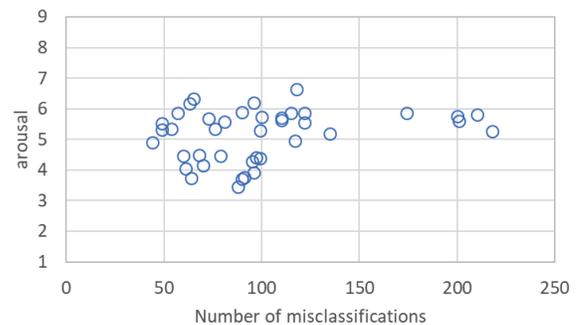}}
	\caption{Relationship between classification performance and subjective scores.}
	\label{fig:emotion-mis}
\end{figure}

It is also observed that the performance significantly differs depending on the video. The number of false classifications of the first five videos occupies 25\% of the entire failure cases. In order to analyze the relationship between the classification performance and video characteristics, Figure \ref{fig:emotion-mis} plots the preference, valence, and arousal scores with respect to the number of misclassifications. The five videos with larger numbers of failure received lower preference scores (2.933 on average) compared with the others (5.518 on average)\footnote{This result is not because of different performance for the two classes. The classification accuracies for the low preference and high preference classes are almost the same, i.e., 0.974 and 0.972, respectively.}. Those videos also have an affective characteristic in common, i.e., low valence and high arousal. While the average valence and arousal scores are 5.254 and 5.157 for all videos, respectively, the videos showing high misclassification rates received the valence and arousal scores of 3.507 and 5.665 on average, respectively. In summary, it seems that the neural activities for videos inducing low preference, low valence, and high arousal are relatively difficult to classify. One possibility that can explain this tendency is the negative emotion (low valence) of stimuli influences on the classification performance. From previous psychological researches, it was revealed that the negative emotion tend to induce more intensive responses than the positive emotion \cite{Peeters90}, and the bias to the negative emotion probably becomes strong with the high arousal in this experiment. Therefore, neural activities related to the negative emotion may overwhelm the QoE-related neural activities so that the classification performance is degraded. 

Moreover, we conducted the same analysis for the score regression for the best case in Table \ref{table:regresult}. We found that the regression for the five videos with high misclassification ratios also shows relatively large errors. These videos are included in the bottom 20\% in terms of the sum of absolute differences between the ground truths and predicted preference scores. 

\section{Conclusion}
\label{sec:conclu}

We have proposed a novel preference prediction approach using EEG based on CNN. We demonstrated significantly improved performance of the proposed method in comparison to previous works for preference prediction. Moreover, we examined various combination of network complexity, feature type, and arrangement of the input matrix.

In the future, other types of brain connectivities will be examined; in particular, an improved version of the phase synchronization may be helpful to enhance the prediction performance. Furthermore, it will be also interesting to consider the characteristics of subjects and videos to obtain the robust performance of QoE recognition.


\section*{Acknowledgment}
This work was supported by Basic Science Research Program through the National Research Foundation of Korea (NRF) funded by the Korea government (MSIT) (NRF-2016R1E1A1A01943283).




\bibliographystyle{IEEEtran}
\bibliography{refs}
%
%
%

\end{document}